\documentstyle[epsf]{article}

\def\abstract#1{\vskip 7mm 
        \begin{center}{\large Abstract}\par \smallskip
                \begin{minipage}[c]{12cm}
                        \small #1
                \end{minipage}
        \end{center}
}
\def\title#1{\begin{center}{\Large\bf #1}\end{center}}
\def\author#1{\vskip 5mm \begin{center}{#1}\end{center}}
\def\address#1{\begin{center}{\it #1}\end{center}}
\def\ug#1#2{{\rm g}^{#1#2}}
\def\dg#1#2{{\rm g}_{#1#2}}
\def\rcp#1{{1\over #1}}

\makeatletter
\@ifundefined{lesssim}{}{}
\@ifundefined{gtrsim}{}{}
\def\vereq#1#2{\lower3pt\vbox{\baselineskip1.5pt \lineskip1.5pt
\ialign{$\m@th#1\hfill##\hfil$\crcr#2\crcr\sim\crcr}}}
\makeatother

\begin{document}

\title{%
  Symmetries of the Gowdy Equations and Spatial Topologies
}
\author{%
  Masayuki TANIMOTO
  \footnote{email: tanimoto@yukawa.kyoto-u.ac.jp}
}
\address{%
  Yukawa Institute for Theoretical Physics,
  Kyoto University, Kyoto 606-8502, Japan \\
  \& \\
  Department of Earth and Space Science, Graduate School of Science, \\
  Osaka University, Toyonaka 560-0043, Japan.
}
\abstract{
  We examine some kinds of discrete symmetries which are
  dynamically preserved, using the (generalized) Gowdy models of the
  first kind.
}


\section{Introduction}

It seems like there are some connections between dynamical properties of
a spacetime and its spatial topology. For example, we know the
recollapsing conjecture \cite{BGT,LW,B}; As is well known, in a positive
curvature (topologically $S^3$) homogeneous and isotropic cosmological
model, the universe contracts and recollapses after an expanding
era. The conjecture claims that this recollapsing property does not
depend on the symmetry imposed on the model, that is, any spacetime
recollapses if the space is {\it topologically} $S^3$ (and if only
appropriate energy and pressure conditions are fulfilled). Thus we may
think of the recollapsing property as a result of the spatial topology.
(For another interesting examples related to asymptotic dynamics, see
\cite{FM,A}.)

In this article we compare some dynamical properties of the
(generalized) Gowdy models of the first kind. The spatial manifold of
this kind is a $T^2$-bundle over $S^1$, and each $T^2$-fiber is
generated by two commuting local Killing vectors. (Such a model was
first considered by Rendall \cite{Ren}, and discussed extensively by the
author \cite{T98,T00}. These models are generalizations of Gowdy's $T^3$
model \cite{Gow}.) There are infinite number of such bundles which are
topologically distinct from each other. However, since each fiber is
generated by (local) Killing vectors in our models the natural reduced
(spatial) manifold is $S^1$ in all cases. Because of this reason we can
compare the dynamical properties of each model in rather detail.

What we investigate concerns symmetries or sets of symmetric data which
are preserved temporarily. More specifically, we consider reflection
symmetries (in a generalized sense). For example, for the usual $T^3$
Gowdy model any set of initial data which are described by even
(spatial) functions evolves in such a way that the data remain even
functions at any time. Also, the same statement is true even if we
replace ``even'' by ``odd'', and in general we will see there is much
larger set of such ``reflection'' symmetries which are preserved. We can
consider corresponding symmetric data for any other models, but the
symmetries are not necessarily preserved. We may interpret this is a
manifestation of the influence of spatial topology.

This article basically treats the same topic as Ref.\cite{T00}, but the
presentation is made in a somewhat different way.

\section{The Gowdy Models of the 1st Kind}

The metric of the Gowdy models can be written in the form
\begin{equation}
  \label{eq:g}
  ds^2=-e^A(d\tau^2-dx^2)+R[e^P(dy+Qdz)^2+e^{-P}dz^2],
\end{equation}
where $A$, $R$, $P$, and $Q$ are functions of $\tau$ and $x$.
There are two commuting Killing vectors for this metric, 
$\partial/\partial y$ and $\partial/\partial z$.
We may think of this metric as that on ${\bf R}^4$, but we can also
compactify the spatial part ${\bf R}^3$ to a $T^2$-bundle over $S^1$ if
imposing appropriate boundary conditions on the metric functions.
In such a case, the model is called a (generalized) Gowdy model of the
1st kind, which we will consider. After the compactification the two
Killing vectors, in general, descend to local Killing vectors,
generating the $T^2$-fibers.

Note that each $T^2$-fiber is characterized by its volume {\it and} two
moduli parameters $(X,Y)\equiv (Q,e^{-P})$.

\section{Symmetries}

As well known \cite{Gow,T00}, the vacuum Einstein equation for metric
function $R$ is given by the simple wave equation
$(\partial_{\tau\tau}-\partial_{xx})R=0$, and we can set without loss of
generality, $R=\tau$. With this choice, the equations for functions $P$
and $Q$ are found to form a closed set of equations, and the remaining
function $A$ is integrable once $P$ and $Q$ are solved. So, we are led
to concentrate on the moduli parameters $X$ and $Y$, and the equations
for them are derivable from the following (reduced) Hamiltonian $H$,
\begin{eqnarray}
  H&=&\int{\cal H} dx, \\
  {\cal H}&=&\rcp2 (\ug
  AB\Pi_A\Pi_B+e^{-2\tau}\dg ABX^A{}'X^B{}'),
\end{eqnarray}
where scripts $A,B,\cdots$ run 1 to 2. We have set
$(X^1,X^2)\equiv(X,Y)$, and $(\Pi_1,\Pi_2)$ is the set of the conjugate
momenta. The ``metric'' $\dg AB$ on the moduli space ${\cal M}$ (spanned
by $X$ and $Y$) is given by the following {\it hyperbolic metric}
\begin{equation}
  \label{eq:hyp}
  dS^2=\frac{dX^2+dY^2}{Y^2}.
\end{equation}
So, the moduli space is also a hyperbolic plane $H^2$ with three
dimensional isometry group Isom$H^2$, of which connected component is
$SL_2{\bf R}$. (Exactly speaking, ${\cal M}$ is not $H^2$ itself, but a
quotient space of it. This point is explained in \S.\ref{sec:4mdls}.)

Note that the data for a moment ($\tau=$ constant), i.e., the
configuration of the spatial manifold, is given by a
loop $l(x)$ in the moduli space:
\begin{equation}
  \label{eq:loop}
  l: [0,1]\longrightarrow{\cal M}, \quad l(0)=l(1),
\end{equation}
and the whole spacetime is described by a one-parameter ($\tau$) family
of such loops: $l(x,\tau)$.

The isometry group Isom$H^2$ provides symmetries of motion, i.e., if
the ``spacetime'' $l(x,\tau)$ is a solution for the vacuum Einstein
equation, then the action $g\cdot l(x,\tau)$ is also another solution
for $g\in{\rm Isom}H^2$. This is one of well-known properties of the
Gowdy equations.

We also should point out a rather trivial but important symmetry, which
is given by the following reparametrization:
\begin{equation}
  \label{eq:repR}
  R\cdot l(x,\tau)= l(-x,\tau).
\end{equation}
As is easily seen, $R$ inverts (or reflects) each spatial configuration
about the origin $x=0$. This also define a symmetry of motion.

\section{The Reflection Operators}

While we may think of the operator $R$ itself as a reflection
operator, it is also possible to generalize reflections using the
isometries Isom$H^2$, i.e., we consider the composition
\begin{equation}
  \label{eq:ref}
  {\cal R}=R\cdot g, \quad g\in{\rm Isom}H^2.
\end{equation}
Any such an operator defines a symmetry of motion. We should
furthermore impose the condition for ${\cal R}$ to form a ${\bf Z}_2$
group (together with the identity). For convenience, defining ${\cal R}$
as a complex function
\begin{equation}
  \label{eq:comp}
   {\cal R}:\;   z= X+i Y\rightarrow{\cal R}(z),
\end{equation}
this condition is expressed by
\begin{equation}
  \label{eq:z2}
  {\cal R}^2(z)=z, \quad \mbox{for all } z\in{\bf C}.
\end{equation}
From this we find two classes of operators; one is given by $R$ itself
\begin{equation}
  \label{eq:RI}
  {\cal R}_I\equiv R,
\end{equation}
and the other is given by a one-parameter family
\begin{equation}
  \label{eq:Rth}
  {\cal R}_\theta\equiv R\cdot\gamma\cdot f_\theta,
  \quad \theta\in[0,\pi),
\end{equation}
where
\begin{eqnarray}
  \label{eq:gam}
  \gamma(z)&=& -\bar z, \\
  \label{eq:fth}
  f_\theta(z)&=& 
  \frac{z\cos 2\theta-\sin2\theta}{z\sin 2\theta+\cos2\theta}.
\end{eqnarray}

However, these operators do not necessarily preserve the boundary
condition imposed on the solution $X(x,\tau)$ and $Y(x,\tau)$, so not
all these operators are appropriate for a given spatial topology.
To account for this point for more detail, we should explain the four
representative models, called the $T^3$, $E^3$, Nil, and Sol models.

\section{The Four Representative Models}
\label{sec:4mdls}

\begin{figure}[hbtp]
  \begin{center}
    \leavevmode
\epsfysize=5cm
\epsfbox{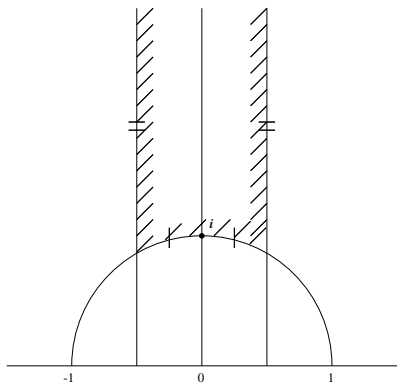}
  \end{center}
\caption{ The fundamental domain ${\cal F}$
  for the moduli space ${\cal M}$ is shown (the shaded region).}
\label{fig:modspc}
\end{figure}

We have noticed any spatial configuration is represented by a loop $l$
on the moduli space ${\cal M}$, but this does not mean the metric
functions $X(x)$ and $Y(x)$ are periodic with respect to the spatial
coordinate $x$. This is because ${\cal M}$ is a quotient space. It is a
classical fact that ${\cal M}$ is represented by $H^2$ {\it modulo}
$SL_2{\bf Z}$, ${\cal M}\simeq H^2/SL_2{\bf Z}$, where $SL_2{\bf Z}$ is
the mapping class group for $T^2$. (This discrete group, $SL_2{\bf Z}$,
is generated by the shift $T:z\rightarrow z+1$ and the inversion
$S:z\rightarrow -1/z$.)  $H^2$ is naturally represented by the upper
half plane ${\bf C}^+\equiv \{ z\in{\bf C}| {\rm Im}(z)>0\}$, and a
fundamental domain ${\cal F}$ for ${\cal M}$ is given by ${\cal F}=
\{ z\in{\bf C}^+| -\frac12 \leq {\rm Re}(z) \leq\frac12, |z|\geq1
\}$. See Fig.\ref{fig:modspc}. For the boundary, the two vertical lines
(at ${\rm Re}(z)=\pm\frac12$) should be identified, and the arc (at
$|z|\leq 1$) is folded about $z=i$ and the opposite sides are
identified, as well. So, the point $z=i$ becomes singular. (These are a
standard fact about the geometric structure on $T^2$. See a standard
text for a detail.) Let us denote a curve on ${\bf C}^+$ which descends
to a loop $l$ on ${\cal M}$, as $\tilde l$. It is $\tilde l$ that has a
direct correspondence to the boundary condition imposed on $X(x)$ and
$Y(x)$.  Apparently, $\tilde l$ is not necessarily a loop, so what is
said above comes. Note that since two (topologically) distinct
$T^2$-bundle over $S^1$ can have the same configuration of moduli
parameters, correspondences of a loop $l$ (or curve $\tilde l$) to a
topology are not unique, but there are not ambiguities between the four
models.

Now, we are in a position to describe the four models. (However, the
following descriptions are not complete enough. See Ref.\cite{T00} for
more complete ones, though they are less pictorial.)

\begin{figure}[hbtp]
  \begin{center}
    \leavevmode
\epsfysize=10.5cm
\epsfbox{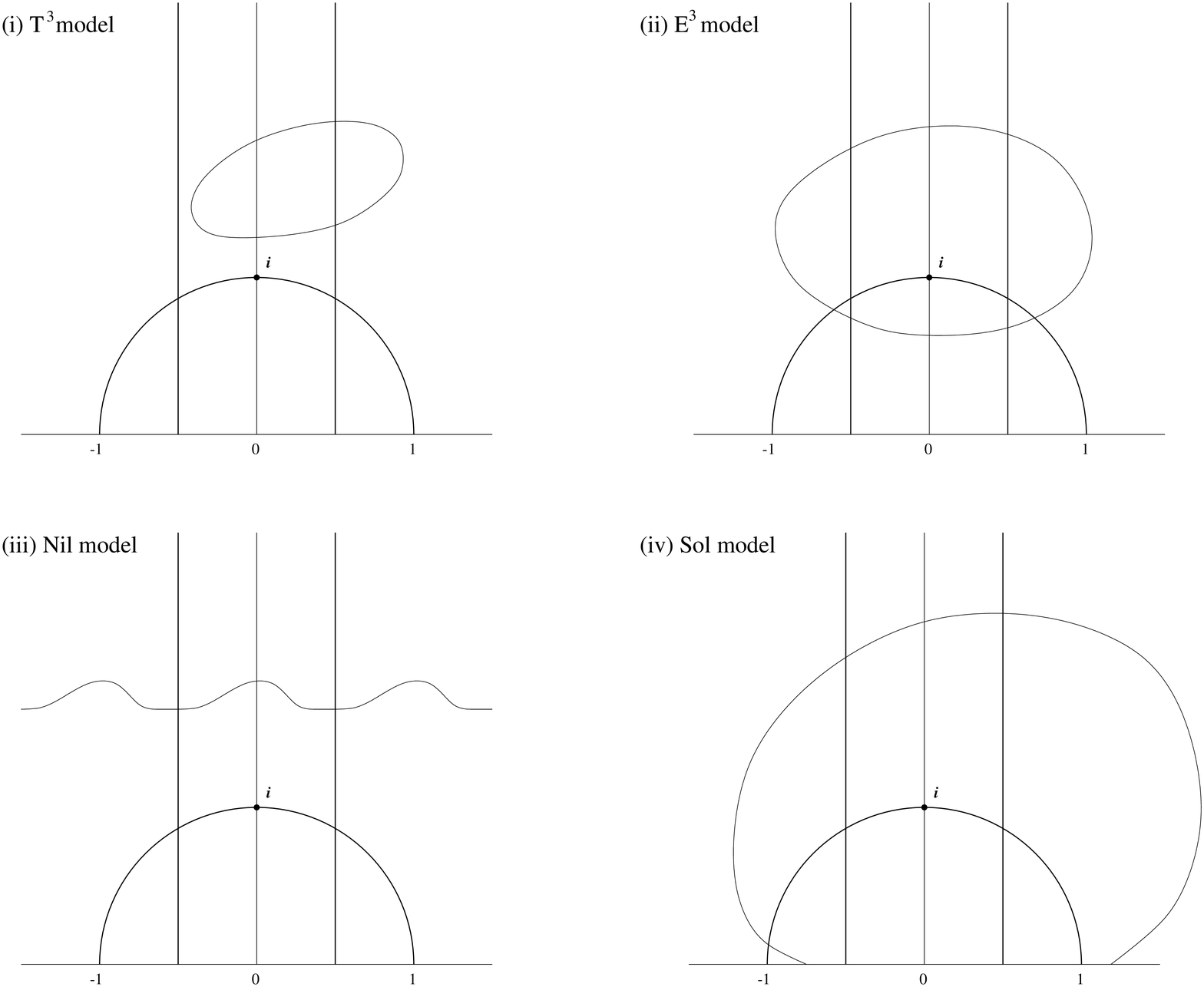}
  \end{center}
\caption{A morphological classification of the four models is shown.
  A possible curve $\tilde l$ is depicted for each model in the upper
  half plane ${\bf C}^+$. Every curve descends to a loop (in ${\cal
  M}$) after identifications generated by $T$ and $S$.}
\label{fig:4mdls}
\end{figure}

{\bf (i) $T^3$ model}: This model is the conventional one, for which the
spatial manifold is topologically $T^3$. The boundary condition for the
metric functions (moduli parameters) is simply given by the periodic
boundary condition.  A loop $l$ for this model is also represented by a
loop $\tilde l$ on ${\bf C}^+$. See Fig.\ref{fig:4mdls} (i). $\tilde l$
does not encircle the singular point $z=i$, so can smoothly contract to
a point, which corresponds to a flat homogeneous space (Bianchi I
space).

{\bf (ii) $E^3$ model}: This model is the representative model for the
models whose spatial part is $T^3$, $T^3/{\bf Z}_2$, $T^3/{\bf Z}_3$,
$T^3/{\bf Z}_4$, or $T^3/{\bf Z}_6$. The boundary condition is periodic,
but the loop $\tilde l$ on ${\bf C}^+$ must encircle the singular point
$z=i$, so {\it cannot} smoothly contract to a point. See
Fig.\ref{fig:4mdls} (ii). The locally homogeneous limit corresponds to a
Bianchi VII${}_0$ space.

{\bf (iii) Nil model}: $\tilde l$ is an open curve which is invariant
under the shifting $T$. One end approaches to ${\rm Re}(z)=+ \infty$,
and the other to ${\rm Re}(z)=- \infty$. See Fig.\ref{fig:4mdls}
(iii). $\tilde l$ descends to a loop on ${\cal M}$ which winds around
the singular point $z=i$ only once. At a locally homogeneous limit,
$\tilde l$ becomes a horizontal straight line, which corresponds to a
Bianchi II space.

{\bf (iv) Sol model}: $\tilde l$ is an open (arc-like) curve for which
both ends approach to ${\rm Im}(z)=0$; one end to $|z|<1$ and the other
end to $|z|>1$. See Fig.\ref{fig:4mdls} (iv). $\tilde l$ descends to a
loop on ${\cal M}$ which winds around the singular point $z=i$ twice or
more.  The locally homogeneous limit corresponds to a Bianchi VI${}_0$
space.

\section{The Invariant Sets}

Table 1 shows the appropriate reflection operators for each model
\cite{T00}. These operators are extracted by requiring they preserve the
appropriate boundary condition for each model, i.e., $\tilde l$
belonging to a model be mapped by one of these operators to another
$\tilde l'$ belonging to the same model.  (More precisely, to get these
reflection operators, we also require a compatibility with a
translation operator \cite{T00}.)

Note that once a reflection operator ${\cal R}$ is given, we can
accordingly define a set of reflection symmetric solutions ${\cal
  S}({\cal R})$:
\begin{equation}
  \label{eq:rss}
  {\cal S}({\cal R})\equiv
  \{l(x,\tau)|\; {\cal R}(l(x,\tau))=l(x,\tau)\}.
\end{equation}
These solutions are invariant under the action of the reflection
operator. We call such a set of solutions an {\it invariant (sub)set}.

\begin{table}[hbtp]
  \begin{center}
\begin{tabular}[c]{|c|c|c|} \hline
  Model & \hspace*{.5em}Reflections\hspace*{.5em} \\ \hline
  ${T^3}$ & ${{\cal R}_I}, \;{{\cal R}_\theta}$
   \\ \hline
  ${E^3}$ & ${{\cal R}_\theta}$
   \\ \hline
  {Nil} & ${{\cal R}_0}$  \\ \hline
  {Sol} & ${{\cal R}_{\pi/4}}$  \\ \hline
\end{tabular}
    \caption{This shows which reflection operators are eligible for each model.
      The parameter $\theta$ in $T^3$ or $E^3$ model can take any value
      in the range $[0,\pi)$, whereas Nil and Sol models can take
      $\theta=0$ and $\pi/4$, respectively. ${{\cal R}_I}$ is possible
      only for $T^3$ model.}
    \label{tab:1}
  \end{center}
\end{table}

Apparently, we have the largest union of such sets for the $T^3$ model,
while that of the Nil or Sol model is the smallest. {\it This is a
  character coming from topological distinctions.}

\section{Dynamical Interpretation}

We interpret the above fact as follows.

Note first that the phase space ${\cal P}$ for the $T^3$ model is given
by the space of the sets of four (smooth) periodic functions
$(X,Y,\Pi_X,\Pi_Y)$. The phase space for Nil or Sol model would of
course be given by another space, since the metric functions for them
are not periodic functions. However, we know it must still be possible
to identify it with ${\cal P}$, since the base spaces of the bundles are
all $S^1$. All we have to do to do this is to re-represent the metric
functions in such a way that they represent deviations from the
configuration of a locally homogeneous limit. Practically, this can be
easily done by replacing the coordinate basis of the metric (\ref{eq:g})
by an invariant basis for the appropriate Bianchi type, which was given
in \S.\ref{sec:4mdls}.

Next consider the reflection operators ${\cal R}_I$ and ${\cal
  R}_\theta$. Although they have been defined so they act on a
configuration space, we can naturally extend them to act on the phase
space using ``Hamiltonian equations'' $\Pi_X=Y^{-2}\dot X$,
$\Pi_Y=Y^{-2}\dot Y$. Then those operators define a set of reflection
symmetric data in ${\cal P}$ by requiring these data be invariant under
the action of one of the operators. These sets are common for all the
models, since the above two (Hamiltonian) equations are common even if
the variables are re-represented in the way described above.

So, we have the same (underlying) phase space and the same reflection
symmetric data sets for all the four models. The Hamiltonians are,
however, different in this setting, and we can compare dynamical
properties of the four models from this point of view. In fact we can
interpret \cite{T00} the result of the previous section as telling that
{\it only the reflection symmetries shown in Table.\ref{tab:1} are
  preserved by the Hamiltonian flow.} This is a manifestation of the
influence of topology to dynamics.

{\bf Remark}: One might think that there were much more possibilities
for another set of reflection symmetric data. However, since we are
interested only in those preserved dynamically, it is sufficient to
consider the operators ${\cal R}_I$ and ${\cal R}_\theta$.

\end{document}